\documentclass[pra,twocolumn,letterpaper,superscriptaddress]{revtex4}
\usepackage{graphicx,amsmath,amssymb,amsfonts,latexsym,color,dcolumn,bm,epsfig,subfigure}
\usepackage[latin1]{inputenc}

\def\bbm[#1]{\mbox{\boldmath $#1$}}

\begin{document}

\title{Three-body amplification of photon heat tunneling}

\author{Riccardo Messina}\email{riccardo.messina@institutoptique.fr}
\affiliation{Laboratoire Charles Fabry, UMR 8501, Institut d'Optique, CNRS, Universit\'{e} Paris-Sud 11, 2, Avenue Augustin Fresnel, 91127 Palaiseau Cedex, France.}

\author{Mauro Antezza}\email{mauro.antezza@univ-montp2.fr}
\affiliation{Universit\'e  Montpellier 2, Laboratoire Charles Coulomb UMR 5221, F-34095, Montpellier, France}
\affiliation{CNRS, Laboratoire Charles Coulomb UMR 5221, F-34095, Montpellier, France}

\author{Philippe Ben-Abdallah}\email{pba@institutoptique.fr}
\affiliation{Laboratoire Charles Fabry, UMR 8501, Institut d'Optique, CNRS, Universit\'{e} Paris-Sud 11, 2, Avenue Augustin Fresnel, 91127 Palaiseau Cedex, France.}

\date{\today}

\begin{abstract}
Resonant tunneling of surface polaritons across a subwavelength vacuum gap between two polar or metallic bodies at different temperatures leads to an almost monochromatic heat transfer which can exceed by several orders of magnitude the far-field upper limit predicted by Planck's blackbody theory. However, despite its strong magnitude, this transfer is very far from the maximum theoretical limit predicted in near-field. Here we propose an amplifier for the photon heat tunneling based on a \emph{passive relay} system intercalated between the two bodies, which is able to partially compensate the intrinsic exponential damping of energy transmission probability thanks to three-body interaction mechanisms. As an immediate corollary, we show that the exalted transfer observed in near-field between two media can be exported at larger separation distances using such a relay. Photon heat tunneling assisted by three-body interactions enables novel applications for thermal management at nanoscale, near-field energy conversion and infrared spectroscopy.
\end{abstract}

\maketitle

\section{Introduction}

Two bodies held at different temperatures and separated by a vacuum gap exchange in permanence heat throughout the thermal electromagnetic field they radiate. At long separation distance (i.e. in far field) this exchange of energy results exclusively from propagative photons emitted by these sources. The blackbody limit given by the famous Stefan-Boltzmann's law sets the maximum heat flux these media can exchange. However, at subwavelength distances (i.e. in near-field regime) the situation radically changes. Indeed, at this scale, evanescent photons which remain confined near the surface of materials are the main contributors to transfer and they participate via tunneling through the vacuum gap \cite{JoulainSurfSciRep05,VolokitinRevModPhys07,KittelPRL05,HuApplPhysLett08,RousseauNaturePhoton09,ShenNanoLetters09,KralikRevSciInstrum11,OttensPRL11}. A significant heat-flux increase results from this transport. In presence of resonant surface modes such as surface plasmons or surface polaritons, collective electron or partial charge oscillations coupled to light waves at the surface, the radiative-heat exchanges can even drastically surpass by several orders of magnitude the prediction of Planck's blackbody theory. This discovery opened new possibilities for the development of innovative technologies for nanoscale thermal management, such as near-field energy conversion (thermophotovoltaic conversion devices \cite{DiMatteoApplPhysLett01,NarayanaswamyApplPhysLett03}), heating assisted data storage (plasmon assisted nanophotolitography \cite{SrituravanichNanoLett04}) or IR sensing and spectroscopy \cite{DeWildeNature06,JonesNanoLetters12}.

Recent theoretical developments \cite{BenAbdallahPRB10,BiehsPRL10} have established that the nanoscale heat transfer between two media separated by a distance $d$ can be revisited using the same concepts as in mesoscopic physics of charge transport. The first step is the expression of the distance-dependent heat flux $\varphi(d)$ under the form of a spectral decomposition $\varphi(d)=\int_0^{+\infty}\frac{d\omega}{2\pi}\phi(\omega,d)$. Hence, the monochromatic near-field heat flux transferred between two media at temperatures $T_1$ and $T_2$ is described in a Landauer-like formalism \cite{BenAbdallahPRB10,BiehsPRL10,LandauerPhilosMag70,ButtikerPRL86} as the sum over all modes
\begin{equation}\phi(\omega,d)=\hbar\omega\,n_{12}(\omega)\sum_p\int\frac{d^2\mathbf{k}}{(2\pi)^2}\mathcal{T}_p(\omega,\mathbf{k},d)\end{equation}
of the energy $\hbar\omega$ of each mode $(\omega,\mathbf{k},p)$ (identified by the frequency $\omega$, the transverse wavevector $\mathbf{k}=(k_x,k_y)$ and the polarization $p$), times a transmission probability $\mathcal{T}_p(\omega,\mathbf{k},d)$ through the separation gap (assuming values between 0 and 1), with $n_{ij}(\omega)=n(\omega,T_i)-n(\omega,T_j)$, $n(\omega,T)=(e^{\hbar\omega/k_BT}-1)^{-1}$ being the distribution function inside the reservoir of modes at temperature $T$. Analysis of arbitrary systems has revealed a rather small efficiency of transfer coefficient for each exchange channel even in the presence of resonant surface modes such as surface polaritons. In this situation the total number of channels (i.e. number of modes) contributing to the heat transfer per unit of surface is set by the position of the cutoff for the transverse component of wavevector $k_c\simeq\log[2/\sqrt{{\rm Im}(\varepsilon_1){\rm Im}(\varepsilon_2)}]/d$ \cite{PendryJPhysCondensMatter99,BiehsPRL10}, where ${\rm Im}(\varepsilon_i)$ denotes the imaginary part of the dielectric permittivity of the material $i$ ($i=1,2$). This allows to estimate two fundamental limits \cite{BenAbdallahPRB10} $\phi_{\rm max}=k_B^2(T_1^2-T_2^2)k_c^2/6\hbar$ and $h_{\rm max}=g_0k_c^2/\pi$ ($g_0=\pi^2k_B^2T/3\hbar$ is the quantum of thermal conductance at temperature $T$) respectively for the monochromatic flux and total heat transfer coefficient between two media. However, these limits are intrinsic to two-body systems but are not universals in general. In the present work we describe heat transfer in three-body systems and we highlight the concept of three-body amplification of heat flux exchanged at nanoscale between two media. In addition we propose a device based on an intermediate \emph{passive relay} which is able to increase the number of coupled modes. This investigation belongs to the vast category of few-body problems whose richness has been largely explored in atomic physics, quantum chemistry and celestial mechanics. Recently, interesting effects have been theoretically discussed in the context of heat transfer \cite{BenAbdallahPRL11,ZhengNanoscale11}.

\section{Results}

\subsection{Physical system and analytical results}

\begin{figure}[h!]
\scalebox{0.3}{\includegraphics{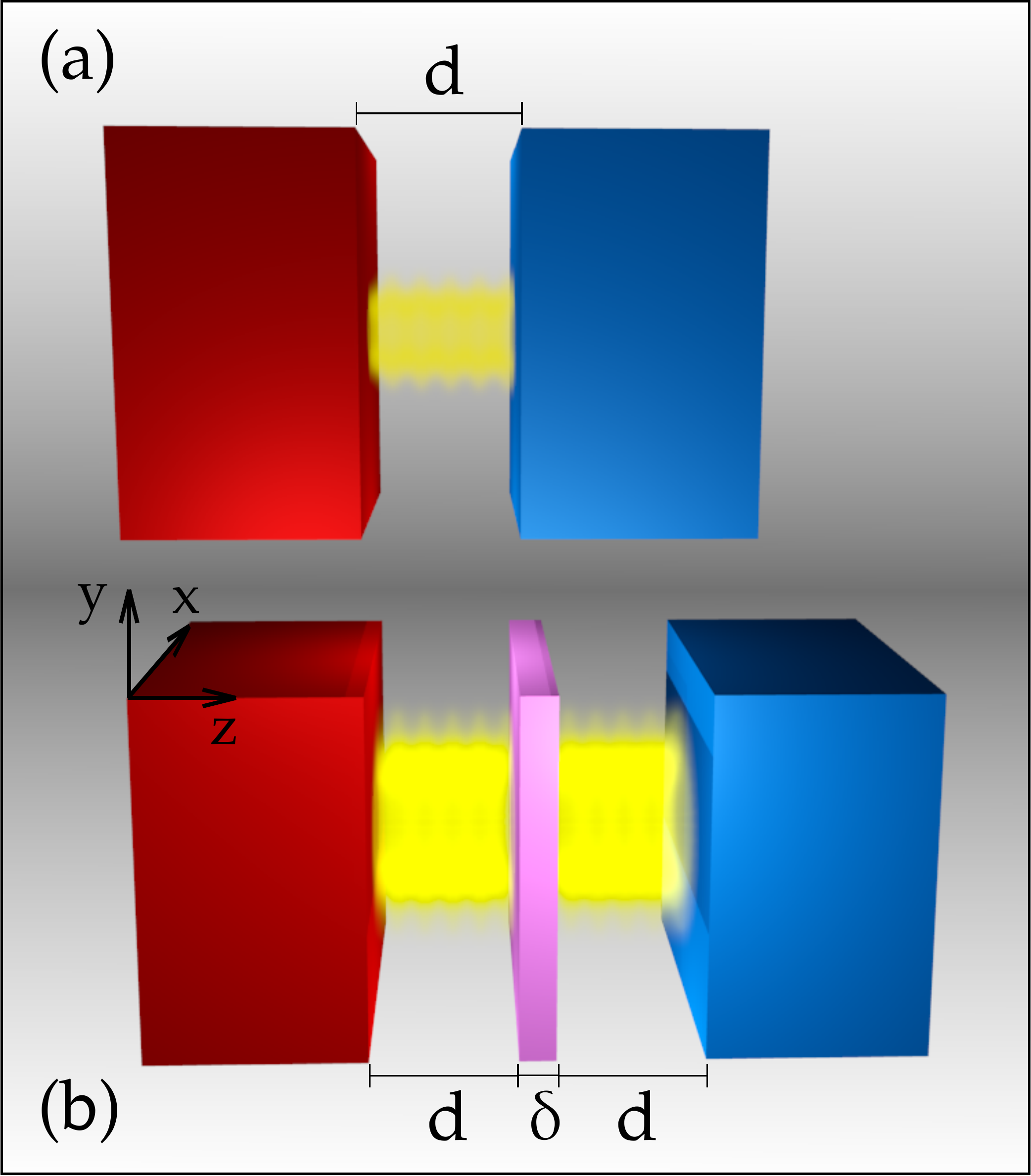}}
\caption{Geometry of the system in the (a) two- and (b) three-slab configurations. The distance $d$ between adjacent slabs (1-2 and 2-3) in the three-body configuration is chosen to be equal to the distance between slabs in the two-body case.}
\label{FigGeometry}\end{figure}

We consider a system consisting of two parallel slabs identified for convenience by the indexes $i=1$ and $i=3$, as shown in figure \ref{FigGeometry}(a).
Each slab has a finite thickness, whereas its transverse extension is much larger than the distances between the slabs, so that the planar surfaces can be considered infinite with respect to the $xy$ plane. We compare this system to a configuration in which a third slab, labeled with $i=2$ and having thickness $\delta$, is placed between slabs 1 and 3 (see figure \ref{FigGeometry}(b)). The system is placed in both cases in a stationary thermodynamical configuration, in which each body is held at a temperature $T_i$ ($i=1,2,3$). Moreover, in order to take into account the radiation impinging on the system from outside, we assume that the two slabs are embedded in an environment described as a source of isotropic thermal radiation at a constant temperature $T_e$, in general different from the temperatures $T_i$. As far as the material properties of the three slabs are concerned, we describe them in terms of three complex dielectric permittivities $\varepsilon_i(\omega)$, meaning that our bodies are characterized here by a non-magnetic local electromagnetic response.

Herein we are interested in comparing the heat flux on body 3 in two- and three-body configurations. The heat-transfer problem in the case of a couple of arbitrary bodies in a thermal environment has been recently solved \cite{MessinaEurophysLett11}. In this case, the heat flux on a body can be expressed as a sum of an evanescent and a propagative contribution: the former depends only on the modes of the electromagnetic field for which the transverse wavevector $\mathbf{k}$ satisfies $ck>\omega$ (i.e. outside the light cone), whilst the latter is defined in the region $ck<\omega$. While the propagative term depends on all the three temperatures involved in the system, the evanescent contribution is only connected to the disequilibrium between the temperatures $T_1$ and $T_3$ of the two bodies. Then, for $T_1\neq T_3$, at distances between them smaller than the thermal wavelength (of the order of some microns at ambient temperature) the evanescent contribution largely dominates on the propagative term. In this near-field regime the monochromatic heat flux on body 3 at frequency $\omega$ can be written under the form of a Landauer expansion
\begin{equation}\label{Phi2s}\phi_{\text{2s}}(\omega,d)=\hbar\omega\,n_{13}(\omega)\sum_p\int_{ck>\omega}\frac{d^2\mathbf{k}}{(2\pi)^2}\mathcal{T}_{\text{2s},p}(\omega,\mathbf{k},d)\end{equation}
where
\begin{equation}\label{T2s}\mathcal{T}_{\text{2s},p}(\omega,\mathbf{k},d)=\frac{4\,{\rm Im}(\rho_{1p}){\rm Im}(\rho_{3p})e^{2ik_zd}}{|1-\rho_{1p}\rho_{3p}e^{2ik_zd}|^2}.\end{equation}
These quantities depend also on the thicknesses of slabs 1 and 3 through the reflection coefficients $\rho_i$ of slab $i$ \cite{MessinaPRA11}.

The scattering procedure developed to investigate heat and momentum transfer between two bodies described in detail in \cite{MessinaPRA11} has been generalized to the case of three bodies in a thermal environment \cite{MessinaPrep}. In this case, as expected on physical grounds, the evanescent
contribution is a function of the three temperatures $T_1$, $T_2$ and $T_3$, while the environmental temperature $T_e$ enters again only in the propagative term. In order to reduce the number of free parameters for the analysis of the amplification mechanism, we will focus our attention of the symmetric case in which the distances between adjacent slabs in the three-body configuration (1\,-\,2 and 2\,-3) are both equal to
$d$, i.e. the distance between slabs 1 and 3 in the two-body configuration (see figure \ref{FigGeometry}). This choice makes $d$ the only relevant distance in both scenarios. Moreover, it makes the minimal distance between any couple of adjacent bodies the same for the two configurations and the optical distance between slabs 1 and 3 \emph{double} in the three-slab case with respect to two slabs. For the three-slab system the near-field expression of the monochromatic heat flux on slab 3 takes the form \cite{MessinaPrep}
\begin{equation}\label{Phi3s}\begin{split}\phi_{\text{3s}}(\omega,d,\delta)&=\hbar\omega\sum_p\int_{ck>\omega}\frac{d^2\mathbf{k}}{(2\pi)^2}\Bigl[n_{12}(\omega)\mathcal{T}^{(12)}_{\text{3s},p}(\omega,\mathbf{k},d,\delta)\\
&\,+n_{23}(\omega)\mathcal{T}^{(23)}_{\text{3s},p}(\omega,\mathbf{k},d,\delta)\Bigr]\end{split}\end{equation}
with
\begin{eqnarray}\label{T3s12}\mathcal{T}^{(12)}_{\text{3s},p}(\omega,\mathbf{k},d,\delta)&=&\frac{4|\tau_{2p}|^2{\rm Im}(\rho_{1p}){\rm Im}(\rho_{3p})e^{2ik_zd}}{|1-\rho_{12p}\rho_{3p}e^{2ik_zd}|^2\,|1-\rho_{1p}\rho_{2p}e^{2ik_zd}|^2}\\
\label{T3s23}\mathcal{T}^{(23)}_{\text{3s},p}(\omega,\mathbf{k},d,\delta)&=&\frac{4\,{\rm Im}(\rho_{12p}){\rm Im}(\rho_{3p})e^{2ik_zd}}{|1-\rho_{12p}\rho_{3p}e^{2ik_zd}|^2}.\end{eqnarray}
Equation (\ref{Phi3s}) shows that the evanescent term splits into two contributions, associated to the couples of temperatures $(T_1,T_2)$ and $(T_2,T_3)$, each one containing a different Landauer transmission probability which is weighted using the corresponding difference $n_{ij}(\omega)$ of thermal populations. In this case, the dependence of the transmission amplitudes on the thickness $\delta$ of the intermediate slab has been explicitly stated. The amplitudes (\ref{T3s12}) and (\ref{T3s23}) are functions of the reflection and transmission amplitudes associated to each slab and to the couple (1,2) treated as a unique body, taking the form
\begin{equation}\rho_{12p}(\mathbf{k},\omega)=\rho_{2p}(\mathbf{k},\omega)+\frac{\tau_{2p}^2(\mathbf{k},\omega)\rho_{1p}(\mathbf{k},\omega)e^{2ik_zd}}{|1-\rho_{1p}(\mathbf{k},\omega)\rho_{2p}(\mathbf{k},\omega)e^{2ik_zd}|^2}.\end{equation}

Let us now consider the choice of temperatures. To avoid the introduction of a supplementary thermal source and thus keep the comparison between the two configurations meaningful we choose for the temperature $T_2$ the value such that the heat flux on body 2 is zero. In the particular case of a quasi-monochromatic flux at frequency $\omega_0$ (which, as we will see, is quite accurate in near-field regime) it can be shown that $T_2$ is solution of the equation $2n(\omega_0,T_2)=n(\omega_0,T_1)+n(\omega_0,T_3)$. We are finally left with the discussion of the material properties of the three slabs. Regarding media 1 and 3, we have chosen two SiC slabs of thickness $5\,\mu$m. As for the intermediate slab, it is a metallic-like medium described by a simple Drude model $\varepsilon(\omega)=1-\omega_P^2/\omega(\omega+i\gamma_P)$, defined by a plasma frequency $\omega_P$ and a relaxation rate $\gamma_P$. As it is well known \cite{EconomouPhysRev69}, this model predicts the existence of a surface mode, a plasmon, at a frequency close to $\omega_P/\sqrt{2}$. Hence, a natural choice for $\omega_P$ to maximize heat transfer is $\omega_P=\sqrt{2}\,\omega_{\rm spp}$, where $\omega_{\rm spp}\simeq1.787\cdot10^{14}\,{\rm rad\,s}^{-1}$ is the surface plasmon-polariton frequency supported by the SiC slabs, matching the SiC and Drude surface-polariton frequencies.

\subsection{Heat-flux amplification}

\begin{figure}[h!]
\scalebox{0.12}{\includegraphics{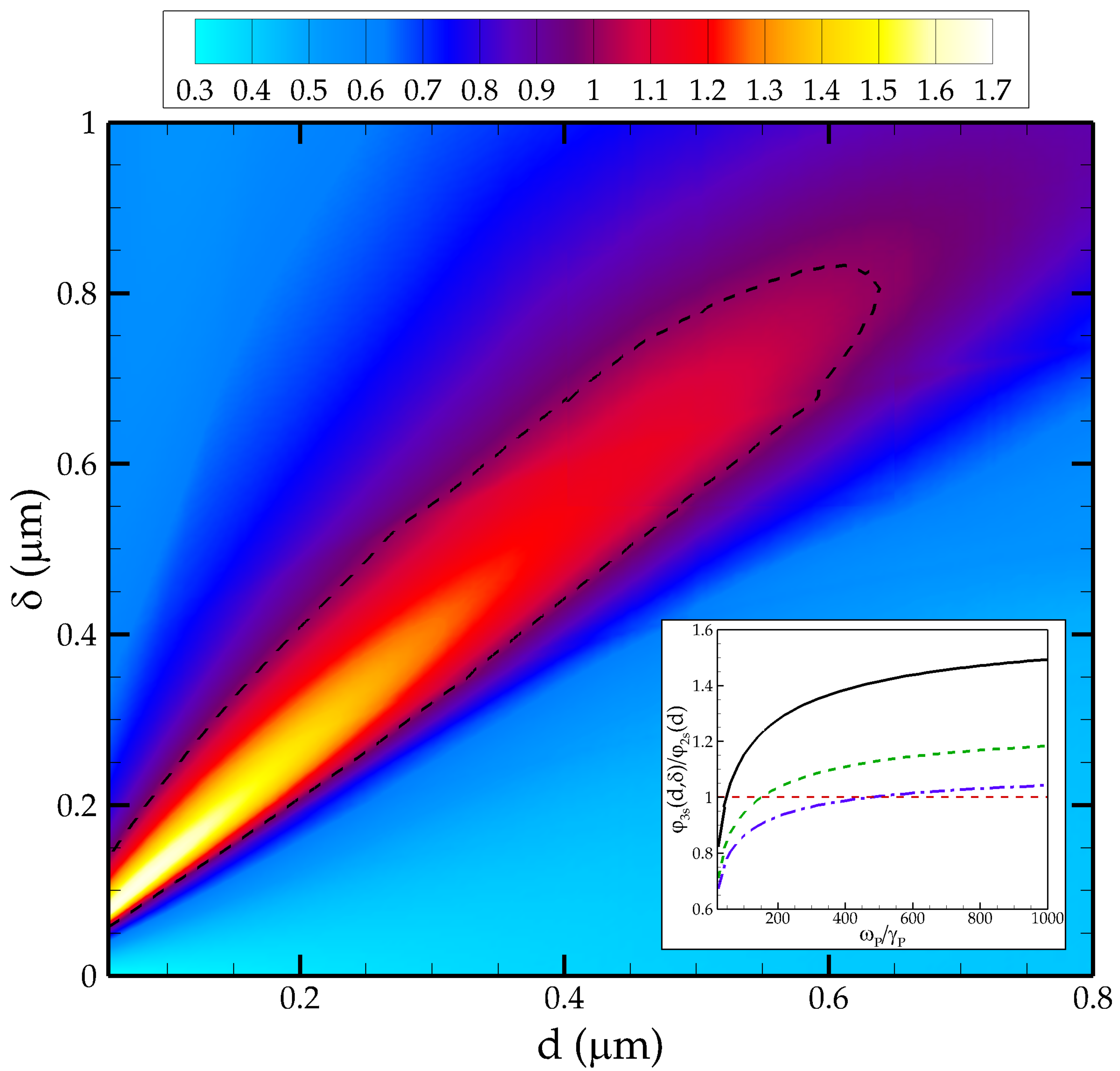}}
\caption{Total heat-flux amplification $\varphi_{\text{3s}}(d,\delta)/\varphi_{\text{2s}}(d)$ as a function of distance and slab thickness. The black dashed line corresponds to $\varphi_{\text{3s}}(d,\delta)/\varphi_{\text{2s}}(d)=1$. The dielectric permittivity of SiC is described using a Drude-Lorentz model \cite{Palik98}, and for slab 2 we have chosen $\gamma_P=10^{-3}\omega_P$. The inset shows the dependence of amplification rates on the relaxation rate $\gamma_P$ for $d=\,200$nm (black solid line), $d=400\,$nm (green dashed line) and $d=600\,$nm (blue dot-dashed line). For each value of $d$ the associated optimal value of $\delta$ has been used.}
\label{Amplification}\end{figure}

We are now ready to calculate the ratio between the total heat flux on slab 3 in the three-slab case with respect to the one in presence of two slabs only. Figure \ref{Amplification} shows the value of this ratio in a range of distances $d\in[50,800]\,{\rm nm}$ and thicknesses $\delta\in[0,1]\,\mu{\rm m}$ of the central slab.
This figure clearly shows the existence of a region of parameters in which the three-body heat flux is larger than its two-body counterpart. As far as the value of the amplification is concerned, in our physical configuration it goes up to 60\% for short distances (around 150\,nm). As shown in the inset of figure \ref{Amplification}, the amplification effect is robust even with respect to the relaxation rate $\gamma_P$ over a range going up to values typical of metals. Moreover, the amplification exists up to distances around 650\,nm and, for each given value of the distance $d$, is limited to a finite interval of $\delta$. More in detail, for a given value of $d$, we see that the amplification assumes a maximum as a function of the thickness of slab 2 for $\delta\simeq d$.

\section{Discussion}

\begin{figure}[h!]
\scalebox{0.12}{\includegraphics{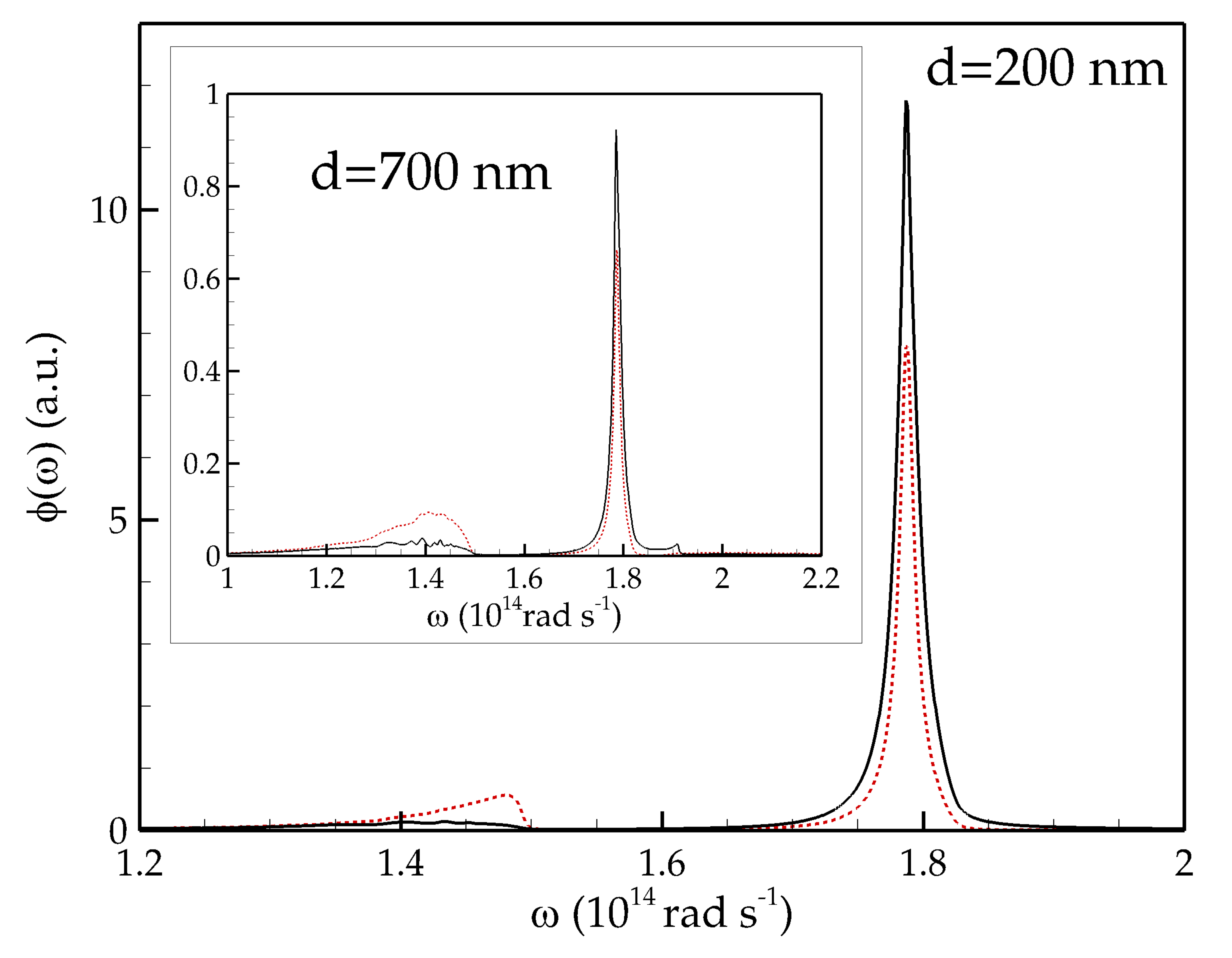}}
\caption{Monochromatic fluxes $\phi_{\text{2s}}(\omega,d)$ (red dashed line) and $\phi_{\text{3s}}(\omega,d,\delta)$ (black solid line) in arbitrary units for two different values of the distance $d$. For each distance the associated optimal value of $\delta$ has been used.}
\label{SpectralFlux}\end{figure}

In order to get an insight on the highlighted amplification mechanism, we first analyze the flux spectra $\phi_{\text{2s}}(\omega,d)$ and $\phi_{\text{3s}}(\omega,d,\delta)$ for different values of the distance $d$ and for the optimal value of the thickness $\delta$ associated to each value of $d$. The results plotted in figure \ref{SpectralFlux} for $d=200\,$nm and $d=700\,$nm show that when the distance between the two slabs approaches zero, the flux tends to become monochromatic at the frequency of the surface polariton. As far as the three-slab term is concerned, we see that in both cases the spectral component of the flux at $\omega_{\rm spp}$ is enhanced with respect to the two-body case. Nevertheless, the presence of slab 2 does not amplify the contribution at smaller frequency, but on the contrary provides a quasi-monochromatic enhancement around the surface polariton frequency $\omega_{\rm spp}$. This explains why, when increasing the distance $d$, the amplification is less important and eventually the ratio between the fluxes goes below one.

\begin{figure}[h!]
\scalebox{0.085}{\includegraphics{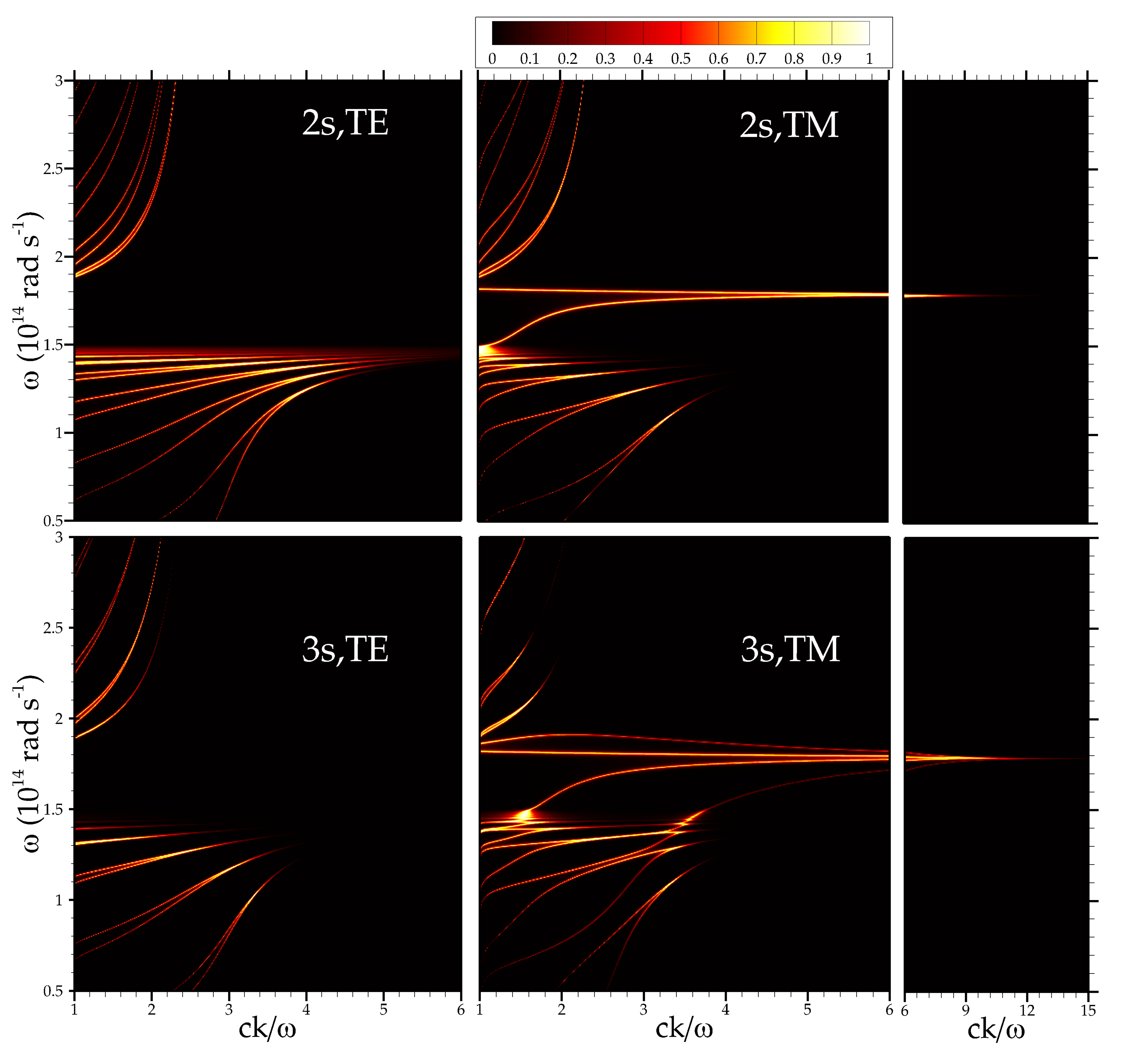}}
\caption{Transmission probabilities $\mathcal{T}_{\text{2s}}$ (upper side) and $\mathcal{T}_{\text{3s}}$ (lower side) for both polarizations as a function of $\omega$ and $k$ at separation distance $d=700\,$nm and for the associated optimal value of $\delta$.}
\label{Modes}\end{figure}

This behaviour can be understood by looking at the coupling efficiencies of modes defined in equations (\ref{T2s}), (\ref{T3s12}) and (\ref{T3s23}). It is important to remark that in the two- and three-body configurations the transmission probabilities are multiplied by different weighting functions $n_{ij}$. In the two-body case, this weight is only given by the difference $n_{13}$, while $n_{12}$ and $n_{23}$ appear in the case of three slabs. However, due to the quasi-monochromaticity of the flux, the temperature $T_2$ is such that $n_{12}(\omega_{\rm spp})=n_{23}(\omega_{\rm spp})=n_{13}(\omega_{\rm spp})/2$. Hence, we compare the transmission probability $\mathcal{T}_{\text{2s}}$ to $\mathcal{T}_{\text{3s}}=(\mathcal{T}_{\text{3s}}^{(12)}+\mathcal{T}_{\text{3s}}^{(23)})/2$. These quantities are plotted in figure \ref{Modes} in the plane $(\omega,k)$ for a given couple $(d,\delta)$. As a first observation, we see that the resonant peak in the spectral flux plotted in figure \ref{SpectralFlux} is due to the presence of resonant surface modes which exist only in TM polarization. We then note that, while in the two-body case we have one symmetric and one antisymmetric surface mode \cite{EconomouPhysRev69}, we observe the appearance of two supplementary modes in presence of an intermediate slab which results from doubling the number of cavities in the system. Moreover, as shown in the right side of figure \ref{Modes}, in the three-slab case these modes remain efficient up to larger values of $k$. These features result in the enhancement and broadening of the resonance peak of the spectral flux. For both polarizations, the presence of supplementary (frustrated) modes at low frequencies explains the secondary peak in the spectral flux in figure \ref{SpectralFlux}. The number and efficiency of those modes is reduced in the three-body configuration: this explains why the amplification mechanism is centered around the resonance frequency of both surface modes (figure \ref{SpectralFlux}).

\begin{figure}[h!]
\scalebox{0.12}{\includegraphics{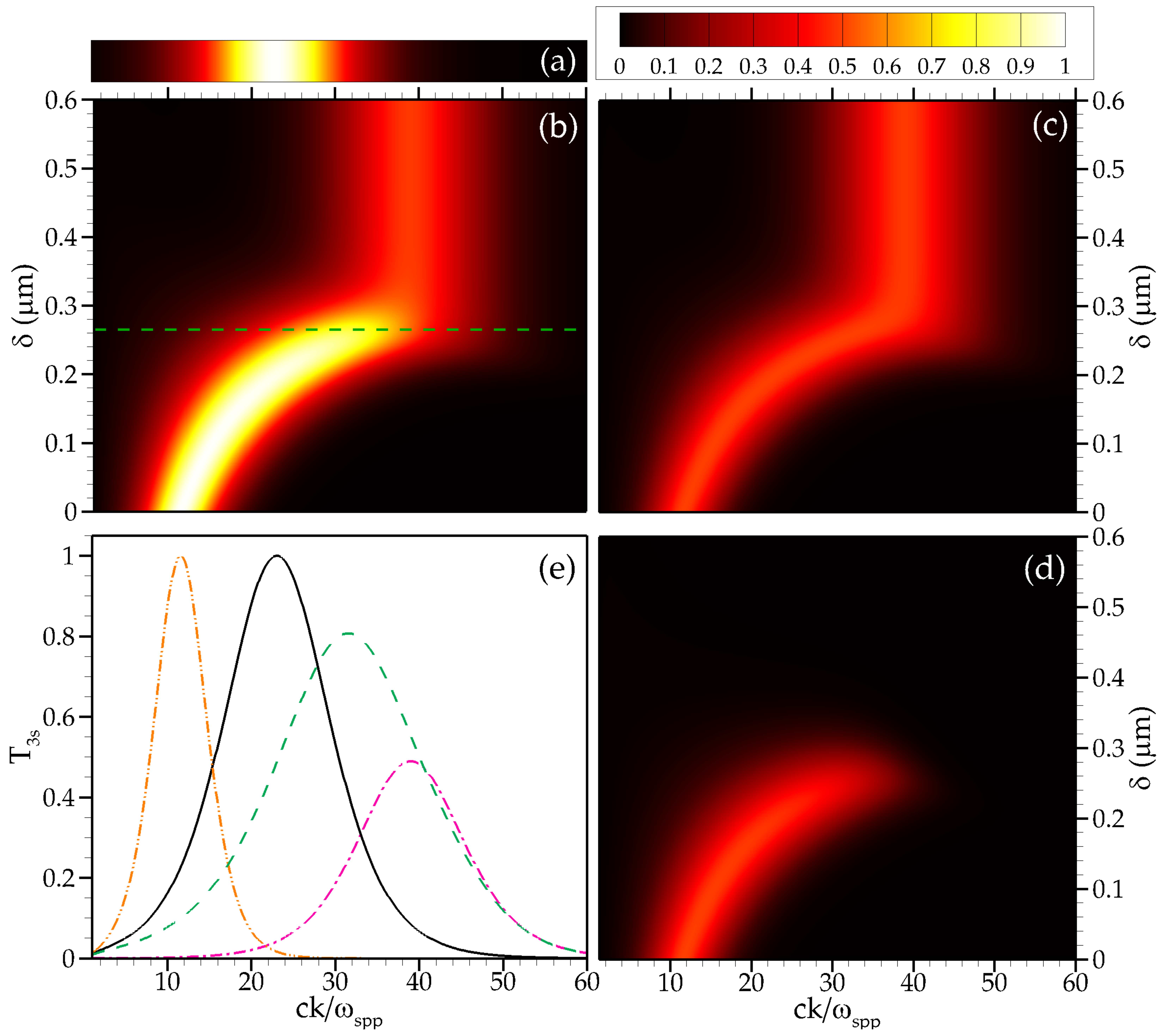}}
\caption{(a) Transmission probability $\mathcal{T}_{\text{3s}}=0.5(\mathcal{T}^{(12)}_{\text{3s}}+\mathcal{T}^{(23)}_{\text{3s}})$ of a three-body SiC\,-\,Drude\,-\,SiC system at the surface plasmon-polariton frequency $\omega_{\rm spp}$ and at a separation distance $d=200\,$nm as a function of dimensionless wavevector $ck/\omega_{\rm spp}$ and thickness $\delta$; (b) $\mathcal{T}_{\text{2s}}$; (c) $0.5\mathcal{T}^{(23)}_{\text{3s}}$; (d) $0.5\mathcal{T}^{(12)}_{\text{3s}}$; (e) Particular case of optimal thickness $\delta=265\,$nm.}
\label{Fk}\end{figure}

To go forward in the physical analysis of the exaltation mechanism we now focus our attention on the transmission probabilities $\mathcal{T}_{\text{2s}}$ and $\mathcal{T}_{\text{3s}}$ (figure \ref{Fk}) at the resonance frequency $\omega_{\rm spp}$ as a function of the wavevector $k$ and thickness $\delta$ for a separation distance $d=200\,$nm. We restrict ourselves to TM polarization, which gives the main contribution to the heat transfer.
In figures \ref{Fk}(a) and \ref{Fk}(b) we show the two- and three-slab transmission probabilities $\mathcal{T}_{\text{2s}}$ and $\mathcal{T}_{\text{3s}}$ respectively. Let us start by discussing in figure \ref{Fk}(b) the two limiting regimes with respect to $\delta$. For $\delta=0$ the system reduces to two SiC slabs separated by a distance $2d$. By comparing this case to the two-slab configuration at distance $d$ in figure \ref{Fk}(a) we observe a shift of the peak and of the cutoff wavevector $k_c$ corresponding to the difference of distance from $d$ to $2d$. Otherwise, when the thickness $\delta$ is larger than $d$, $\mathcal{T}_{\text{3s}}$ becomes independent on $\delta$, corresponding to the fact that the slab 2 is seen as semi-infinite. In this limit, figure \ref{Fk}(b) shows that the replacement of the SiC slab 1 with a semi-infinite slab described by the Drude model produces a shift of $k_c$ toward larger values. This is consistent with the fact that $k_c$ is an increasing function of $1/\sqrt{{\rm Im}(\varepsilon_1){\rm Im}(\varepsilon_2)}$ and that at the plasmon frequency $\omega_{\rm spp}$ the imaginary part of the Drude dielectric permittivity is smaller than the one of the Drude-Lorentz model describing the SiC. Despite this shift, the integral over $k$ (i.e. the monochromatic flux) is still smaller than in the case of two slabs at distance $d$. This is a direct consequence of the fact that $T_2<T_1$.

Hence, we can say that the cases $\delta=0$ and $\delta\gg d$ are as a matter of fact two-body configurations (SiC\,-\,SiC and Drude\,-\,SiC respectively). The intermediate region, corresponding to values of $\delta$ around $d$, results from a purely three-body effects in which the results is intimately connected to the presence of both bodies 1 and 2. A deeper understanding of this transition comes from figures \ref{Fk}(c) and \ref{Fk}(d), representing separately the two contributions to the three-slab configuration $0.5\mathcal{T}^{(23)}_{\text{3s}}$ and $0.5\mathcal{T}^{(12)}_{\text{3s}}$ respectively. The former has exactly the structure of a two-body transmission amplitude, as evident from the comparison between equations (\ref{T2s}) and (\ref{T3s23}). It corresponds to the exchange between the couple (1,2) treated as a unique body at temperature $T_2$ and the body 3. The plot of this coefficient in figure \ref{Fk}(c) shows that this term can be associated to the transition, as a function of $\delta$, from body 2 to body 1.
As a matter of fact, the peak of the transmission amplitude is for any thickness approximately 0.5, but the range of wavevectors (i.e. the number of modes) contributing to the effect moves from the one associated to a Drude material at distance $d$ to the one corresponding to a SiC slab at distance $2d$. On the contrary, the contribution $\mathcal{T}^{(12)}_{\text{3s}}$ can be thought as an exchange between bodies 1 and 3 mediated by the presence of body 2. The presence of the intermediate slab is manifest both in the temperature-dependent term $n_{12}(\omega)$ and in the fact that the product of imaginary part of reflection coefficients $\rho_1$ and $\rho_3$ is weighted over a coefficient dependent on body 2 and in particular proportional to the square modulus of its transmission coefficient. This explains why this term starts contributing below a given value of $\delta$ (of the order of 300\,nm in figure), at which the slab 2 is no longer seen as semi-infinite. It is in proximity of this value that the system exploits the shift of the cutoff wavevector due to the replacement of the SiC with the Drude material and at the same time the contribution for smaller wavevectors guaranteed by the presence of slab 1. This is even more evident in figure \ref{Fk}(e). In this case we represent the two-slab configuration at distance $d$, and the three-slab cases for $\delta=0$, $\delta\gg d$ and $\delta=265\,$nm, representing in this case the optimal value. The curve corresponding to this value of $\delta$ shares the behaviour for $k\to\infty$ with the case $\delta\gg d$, but differs from this curve at smaller values of $k$.

In conclusion, we have proposed and characterized a new passive amplifier based on a three-body assisted tunneling mechanism. By remarkably enhancing near-field exchanges with respect to a two-body system, it allows at the same time to increase the magnitude of heat transfer and to double the separation distance, without any additional source of energy. Due to its quasi-monochromaticity and filtering effect, this mechanism could also be exploited to improve considerably the efficiency of near-field energy conversion devices by increasing the photocurrent generation in thermophotovoltaic cells, as well as reducing the heating of the cell. This work also paves the way to the study of near-field heat transport in complex plasmonic systems mediated by many-body interactions at mesoscopic scale.

\begin{acknowledgments}
The authors thank S.-A. Biehs and D. Felbacq for fruitful discussions. M. A. acknowledges financial support from the Julian Schwinger Foundation. P. B.-A. acknowledges the support of the Agence Nationale de la Recherche through the Source-TPV project ANR 2010 BLANC 0928 01.
\end{acknowledgments}

\end{document}